\documentclass{kluwer}
\usepackage{graphicx,epsfig,amssymb}
\begin{document}
\begin{article}
\begin{opening}
\title{
On the connection between the Nekhoroshev theorem and Arnold Diffusion}
\author{Christos \surname{Efthymiopoulos}}
\runningauthor{C. Efthymiopoulos}
\runningtitle{Nekhoroshev theorem and Arnold diffusion}
\institute{Research Center for Astronomy and Applied Mathematics,
Academy of Athens, Soranou Efessiou 4, 115 27 Athens, Greece}
\begin{abstract}
The analytical techniques of the Nekhoroshev theorem are used to provide
estimates on the coefficient of Arnold diffusion along a particular resonance
in the Hamiltonian model of Froeschl\'{e} et al. (2000). A resonant normal
form is constructed by a computer program and the size of its remainder
$||R_{opt}||$ at the optimal order of normalization is calculated
as a function of the small parameter $\epsilon$. We find that
the diffusion coefficient scales as $D\propto||R_{opt}||^3$,
while the size of the optimal remainder scales as $||R_{opt}||
\propto\exp(1/\epsilon^{0.21})$ in the range $10^{-4}\leq\epsilon
\leq 10^{-2}$. A comparison is made with the numerical results of
Lega et al. (2003) in the same model.
\end{abstract}
\end{opening}

%%%%%%%%%%%%%%%%%%%%%%%%%%%%%%%%%%%%%%%%%%%%%%%%%%%%%%%%%%%%%%%%%%%%%%%
\section{Introduction}
%%%%%%%%%%%%%%%%%%%%%%%%%%%%%%%%%%%%%%%%%%%%%%%%%%%%%%%%%%%%%%%%%%%%%%%
In a series of instructive papers, Froeschl\'{e} et al. (2000, 2005),
Lega et al. (2003), and Guzzo et al. (2005) presented the results of
a detailed numerical investigation of the phenomenon of {\it Arnold diffusion}
(Arnold 1964) in a Hamiltonian system of three degrees of freedom that
satisfies sufficient conditions for the holding of the {\it Nekhoroshev
theorem} (Nekhoroshev 1977, Benettin et al. 1985, Lochak 1992, P\"{o}shel
1993). The aim of these studies was to establish quantitative estimates as
regards a) the critical value of the small parameter $\epsilon_c$ below which
the system enters into the so-called `Nekhoroshev regime', and b) the dependence
of the local diffusion coefficient $D$, along a particular resonance, on
$\epsilon$. In Guzzo et al. (2005) the authors reported to have also
numerically observed global diffusion over an extended domain of the
Arnold web. Other numerical studies over the years related to the same
problem are: Kaneko and Konishi (1989), Wood et al. (1990), Dumas and Laskar
(1993), Laskar (1993), Skokos et al. (1997), Giordano and Cincotta (2004).
Some early results of the literature are briefly commented in the
discussion.

In what follows we focus on an analysis of point (b) above.
According to Lega et al. (2003), a tedious numerical calculation
yields that the local diffusion coefficient $D$ along a resonance
depends monotonically on $\epsilon$. While the limited range of the
numerical data did not allow for a precise fitting, the authors
presented evidence that the numerical function $D(\epsilon)$ was
decreasing faster than a power-law $D(\epsilon) \propto \epsilon^b$,
while it was consistent with an exponential law $D(\epsilon)\propto
\exp(-1/\epsilon^a)$ for some unspecified constant $a$. A subsequent
study in a mapping model analogous to the above Hamiltonian model
yielded the concrete estimate $a\simeq 0.28$ (Froeschl\'{e} et al.
2005).

In the present paper our aim is to provide a deeper understanding of
these numerical results by having recourse, precisely, to the
analytical apparatus of the Nekhoroshev theory, that is, the
construction of a {\it resonant Birkhoff normal form} and of
estimates on the size of the remainder $R$ of the normal form at the
optimal order of normalization. The optimal remainder function
$R_{opt}$ is a crucial quantity for the dynamics because it
represents the true size of the perturbation of the system that
corresponds to its deviation from an integrable system. Indeed, in
Nekhoroshev theory all bounds on the variations of the actions
follow from estimates on the size of $R_{opt}$. In order, therefore,
to establish the link between the Nekhoroshev theory, on the one
hand, and the problem of the speed of diffusion, on the other hand,
in the present paper we seek to determine:

a) the dependence of the local diffusion coefficient $D$ on the size
of the remainder $||R_{opt}||$ at the optimal order of normalization,
and

b) the precise dependence of $||R_{opt}||$ on $\epsilon$. In
particular, we seek to calculate $||R_{opt}||$ when $\epsilon$
becomes small enough so that the exponentially small scaling of
$||R_{opt}||$ with $\epsilon$ clearly shows up. This required to
consider values of $\epsilon$ one order of magnitude smaller than
the values considered in the numerical experiments of Lega et al.
(2003).

We can immediately state a brief summary of our main results. We find that:

a) $D$ scales with $||R_{opt}||$ as a power law $D\propto ||R_{opt}||^c$,
with $c\simeq 3$. To determine this relation, use was made of the data of
Lega et al. (2003) for the numerical values of $D$, and a program was written
by the author in order to compute the Birkhoff normal form and thereby the
remainder $R_{opt}$.

b) In the range $10^{-4}\leq \epsilon\leq 10^{-2}$, $||R_{opt}||$
scales as $||R_{opt}||= \exp(-b/\epsilon^a)$ with $a\simeq 0.21$ and
$b\simeq 3.22$. This also yields the exponential scaling $D\propto
\exp(-3b/\epsilon^a)$. Note that the exponentially small scaling
shows up clearly only for $\epsilon\leq 0.001$, while in the range
$0.001<\epsilon\leq 0.02$ a power-law fitting
$||R_{opt}||\propto\epsilon^{2.45}$ yields marginally better results
than the exponential fitting. In fact, we find that although the
onset of the `Nekhoroshev regime' can be placed at a threshold value
$\epsilon\approx 0.01$, the deviation of the remainder from a power
law becomes clear only for $\epsilon$ one order of magnitude smaller
than this value.

The computational algorithm used to obtain the above results is described
in section 2. It follows essentially the normalization scheme described
in the lecture notes on exponential stability of Giorgilli (2002, pp.86-87).
This scheme is different from the traditional scheme based on normalization
by successive orders of $\epsilon$. We further modify it to avoid using
$K-$truncated resonant modules (see section 2). People working on mathematical
aspects of the Nekhoroshev theory are definitely familiar with all related
notions. However, a somewhat extended description is given in section 2,
to help rendering the subject more accessible also to people oriented
towards the applications.

Section 3 then focuses on the implementation of the algorithm in the particular
model of Froeschl\'{e} et al. (2000). By constructing the normal form and
identifying the optimal order of normalization, the size of $R_{opt}$ is
evaluated for various values of $\epsilon$. This yields the exponential
scaling of $||R_{opt}||$ on $\epsilon$. Using also the data of Lega et al.
(2003) yields the power-law relation of $D$ with $||R_{opt}||$.
Section 4 contains a brief discussion of the results.

%%%%%%%%%%%%%%%%%%%%%%%%%%%%%%%%%%%%%%%%%%%%%%%%%%%%%%%%%%%%%%%
\section{The normal form algorithm}
%%%%%%%%%%%%%%%%%%%%%%%%%%%%%%%%%%%%%%%%%%%%%%%%%%%%%%%%%%%%%%%

%------------------------------------------------------------------------
\subsection{The classical construction}
%-----------------------------------------------------------------------
The analytical part of the Nekhoroshev theorem is based on the construction
of a normal form for a $n$ degrees of freedom Hamiltonian system of the form
\begin{equation}\label{ham}
H(J,\phi)=H_0(J) + \epsilon H_1(J,\phi)
\end{equation}
where $(J,\phi)\equiv(J_1,J_2,...,J_n,\phi_1,\phi_2,...\phi_n)$ are
action-angle variables, $H_0$ satisfies appropriate non-degeneracy
and convexity conditions, and $H$ is analytic in a complexified
domain of the actions and the angles. We shall be concerned with a
{\it local} construction of a normal form for the Hamiltonian
(\ref{ham}), valid within some (small) preselected open domain
${\cal W}_a$ of the action space (the index $a$ means action space).
The analyticity condition implies that if $H_1$ is Fourier expanded
\begin{equation}\label{fourh1}
H_1(J,\phi)=\sum_{k\in\mathbb{Z}^n}H_{1,k}(J)\exp(ik\cdot\phi)\big)
\end{equation}
there are positive constants $A,\sigma$ such that for all values of
$J\in{\cal W}_a$ the coefficients $H_{1,k}$ are bounded by
\begin{equation}\label{expbd}
|H_{1,k}(J)|\leq A\exp(-|k|\sigma)~~.
\end{equation}
The function $H_1$ is analytic in the domain ${\cal W}_a\times
\mathbb{T_\sigma}^n$, where $\mathbb{T_\sigma}^n\equiv\{\phi:
\mbox{Re}{\phi}\in\mathbb{T}^n, |\mbox{Im}{\phi_i}|<\sigma,
i=1,\ldots n\}$. Furthermore, bounds on the function $H_1$ can be
found in domains ${\cal W}_a\times \mathbb{T_{\sigma'}}^n$, for any
positive $\sigma'<\sigma$.

Now, the purpose of the normal form construction is to perform a
series of canonical transformations $(J,\phi)\rightarrow
(J^{(1)},\phi^{(1)}), \rightarrow (J^{(2)},\phi^{(2)}), \ldots$,
such that, after $r$ normalization steps, the Hamiltonian is
expressed as
\begin{eqnarray}\label{hamnorm}
H^{(r)}(J^{(r)},\phi^{(r)})&\equiv
&H(J(J^{(r)},\phi^{(r)})\phi(J^{(r)},\phi^{(r)})) \nonumber\\
&=&Z^{(r)}(J^{(r)},\phi^{(r)})+R^{(r)}(J^{(r)},\phi^{(r)})~~.
\end{eqnarray}
In the last expression, the first term
$Z^{(r)}(J^{(r)},\phi^{(r)})$, called the {\it normal form}, is
constructed in such a way as to yield a simple dynamics in the
variables $(J^{(r)}, \phi^{(r)})$ (e.g. in the non-resonant or
simply resonant case the Hamiltonian $Z^{(r)}(J^{(r)},\phi^{(r)})$
is integrable). On the other hand, the second term
$R^{(r)}(J^{(r)},\phi^{(r)})$, called the {\it remainder},
represents a perturbation to the dynamics induced by the normal
form. The goal of the theory is to proceed with the normalization as
far as necessary in order to minimize the size of the remainder. In
exceptional cases of {\it integrable} Hamiltonian systems one may
proceed by infinitely many steps, within some domain of convergence,
and show that the remainder tends to zero in the limit
$r\rightarrow\infty$. In the generic case of non-integrable systems,
however, one can only reduce $R^{(r)}(J^{(r)},\phi^{(r)})$ to a
finite minimum size. This is found at a specific order of
normalization which is hereafter called the optimal order of
normalization $r_{opt}$. The size of the remainder $R^{(r)}(J^{(r)},
\phi^{(r)})$ for any other order $r\neq r_{opt}$ is bigger than the
size of $R^{(r_{opt})}(J^{(r)},\phi^{(r)})$. Standard theory yields
estimates $r_{opt}\sim 1/\epsilon^b$ and $||R^{(r_{opt})}||\sim
\exp(-1/\epsilon^a)$ for some positive integers $a,b$ depending on
the number of degrees of freedom.

The classical construction leading to expressions of the form (\ref{hamnorm})
is nowadays analyzed in many books (see e.g. Boccaletti and Pucacco (1996),
Contopoulos (2002), Morbidelli (2002), or Ferraz-Mello (2007) for instructive
introductions). The main steps can be summarized as follows:

a) One makes a choice as regards which Fourier terms
$H_{1,k}(J)\exp(ik\cdot\phi)$ in $H_1$ will be retained and which
terms will be eliminated in the normalized Hamiltonian. The terms to
be retained are specified by their integer vector $k$ belonging to a
particular subset of $\mathbb{Z}^n$ called {\it resonant module}
(denoted hereafter by ${\cal M}$)). The choice of resonant module is
dictated by the location of the domain ${\cal W}_a$ in the action
space, i.e., by whether this domain is close to or far from
particular resonances (see subsection 2.3).

b) The Fourier terms not belonging to $\cal{M}$ are the ones to be
eliminated by a canonical transformation. Let us consider the first
normalization step. We shall consider canonical transformations of
old variables $(J,\phi)$ to new variables $(J^{(1)},\phi^{(1)})$
obtained via a Lie generating function $\chi_1(J^{(1)},\phi^{(1)})$.
The function $\chi_1(J^{(1)},\phi^{(1)})$ will be selected so as to
be of the `first order of smallness' (the precise meaning of this
sentence is analyzed below). The canonical transformation is defined
by
\begin{equation}\label{lietran}
J=\exp(L_{\chi_1})J^{(1)},~~~
\phi=\exp(L_{\chi_1})\phi^{(1)}
\end{equation}
where $L_{\chi_1}\equiv\{\cdot,\chi_1\}$ is the Poisson bracket
operator, and $\exp(L_{\chi_1})$ is the formal exponential
development of $L_{\chi_1}$. A well known property of Lie
transformations is that for any function of the canonical variables
$f(q,p)$ one has
\begin{equation}\label{liecan}
\exp(L_\chi)f(q,p)=f(\exp(L_\chi)q,\exp(L_\chi)p)~~.
\end{equation}
As a result, the transformed Hamiltonian, after the action of the generating
function $\chi_1$ is given by
\begin{equation}\label{ham1}
H^{(1)}=\exp(L_{\chi_1})H =
H + L_{\chi_1}H + {1\over 2}L_{\chi_1}^2H+\ldots
\end{equation}

In order to specify the function $\chi_1(J^{(1)},\phi^{(1)})$, we
select from $H_1$ all the terms satisfying the following two
conditions: a) being of the `first order of smallness', and b) not
belonging to ${\cal M}$. Denoting by $h_1$ the sum of these terms,
the terms are eliminated by selecting $\chi_1$ so as to satisfy:
\begin{equation}\label{homo1}
L_{\chi_1}H_0 + h_1=0~~.
\end{equation}
Equation (\ref{homo1}) is called homological. It is  reduced to a
diagonal system of linear algebraic equations. Namely, writing $h_1$ as
$$
h_1=\sum_kh_{1,k}(J)\exp(ik\cdot\phi)
$$
and setting
$$
\chi_1=\sum_k\chi_{1,k}(J)\exp(ik\cdot\phi)
$$
we find, through (\ref{homo1}), the solution for the coefficients
$\chi_{1,k}(J)$ given by:
\begin{equation}\label{homosol1}
\chi_{1,k}(J)={h_{1,k}(J)\over i k\cdot\omega(J)}
\end{equation}
where
\begin{equation}\label{ome}
\omega(J)=\nabla_JH_0(J)
\end{equation}
is the frequency vector of the unperturbed Hamiltonian at the point $J$ of
the action space.

c) In the same way we eliminate the terms of the second order of
smallness in $H^{(1)}$ or, in general, the terms of the r-th order
of smallness in $H^{(r-1)}$. The recursive formulae for the r-th
canonical transformation $J^{(r-1)}=\exp(L_{\chi_r})J^{(r)}$,
$\phi^{(r-1)}=\exp(L_{\chi_r})\phi^{(r)}$ are:
\begin{eqnarray}\label{nfrec}
~&~&\{H_0,\chi_r\}+h^{(r-1)}_r=0\nonumber\\
~&~&H^{(r)}=\exp(L_{\chi_r})H^{(r-1)}~~.
\end{eqnarray}
Following iteratively the above procedure, the function $H^{(r)}$ takes the
form (\ref{hamnorm}). The remainder consists of all the terms of $H^{(r)}$
of `order of smallness' larger than $r$. It can be written as
\begin{equation}\label{remser}
R^{(r)}=h^{(r)}_{r+1}+h^{(r)}_{r+2}+\ldots
\end{equation}
This series is an analytic function in a restriction of the domain of
analyticity of the original hamiltonian provided that the Lie transformation
in (\ref{nfrec}) is convergent at every step (which is ensured for `small
enough' generating functions $\chi_r$).

%------------------------------------------------
\subsection{Book-keeping}
%------------------------------------------------
A clarification of the meaning of `order of smallness' is now in
order. When one implements the recursive scheme of
Eqs.(\ref{lietran}) and (\ref{nfrec}), one needs to decide how
should $H^{(r)}$ at every step be split into terms of different
orders. Such a splitting should be based on the size of each term
relative to the size of the other terms. In a computer program, it
is customary to introduce a so-called `book-keeping' parameter
$\lambda$. This is a parameter the powers of which appear as
coefficients in front of the various terms in the expansion. A term
with coefficient $\lambda^p$ is said to be `of order of smallness'
$p$. This helps separating the terms in the program by, say,
recalling the $\mbox{Coefficient}(h^{(r-1)},\lambda,r)$ in an
algebraic program like Mathematica in order to get the term
$h^{(r-1)}_r$ in Eq.(\ref{nfrec}). As the program carries on the
normalization, the book-keeping coefficients change according to the
rule that the coefficient of the Poisson bracket of two terms with
coefficients $\lambda^p$ and $\lambda^q$ is $\lambda^{p+q}$. In the
end we set $\lambda=1$ to recover the correct values of all the
coefficients.

This way of organizing the series is practical, but it also reflects
a fundamental decision on what quantity we consider to represent
`smallness' in the perturbation series. If the overall size of $H_1$
is an $O(1)$ quantity, then it is customary to identify $\epsilon$
itself as the small parameter. In that case the normalization scheme
proceeds by ascending powers of $\epsilon$, and the generating
functions $\chi_r$ are of order $O(\lambda^r)\equiv O(\epsilon^r)$.
We shall see in subsection 2.4 how to modify the `book-keeping' so
as to separate terms of different smallness which co-exist within
$H_1$, and then within $H^{(r)}$, provided that the function $H_1$
satisfies analyticity conditions of the form (\ref{expbd}).

%------------------------------
\subsection{Resonant K-module}
%------------------------------
The choice of resonant module depends on the location of the domain
${\cal W}_a$ in the action space and on the properties of the
unperturbed Hamiltonian $H_0$. Let us consider the homological
equation of the first order under the choice of book-keeping
$\lambda\equiv\epsilon$. This reads:
\begin{equation}\label{homo2}
\{H_0,\chi_1\} = -\epsilon\tilde{H}_1
\end{equation}
where $\tilde{H}_1$ is the part of $H_1$ containing terms to be eliminated.
The solution of (\ref{homo2}) is
\begin{equation}\label{homosol2}
\chi_1=\epsilon\sum_{k\in\mathbb{Z}^n/\cal{M}}{H_{1,k}(J)\exp(ik\cdot\phi)\over
ik\cdot\omega(J)}~~.
\end{equation}
Provided that the resonant manifolds $k\cdot\omega(J)=0$ are dense
in the action space, for any small open domain ${\cal W}_a$ there
are infinitely many resonant manifolds crossing ${\cal W}_a$. This
implies that there is a dense set of values of the actions $J$
within ${\cal W}_a$ for which a divisor in (\ref{homosol2}) becomes
exactly equal to zero. In practice, this means that a generating
function like $\chi_1$ in (\ref{homosol2}) cannot in fact be
determined in any open domain ${\cal W}_a$. In Nekhoroshev theory,
an often stated remedy to this problem consists of splitting the set
of Fourier harmonics $\exp(ik\cdot\phi)$ into low order and high
order harmonics according to whether $|k|\equiv |k_1|+\ldots+|k_n|$
is smaller or larger than some positive integer $K$ (see, for
example, Morbidelli and Guzzo 1997). The choice of value of $K$ is
arbitrary, but the choice $K\sim 1/\epsilon^a$, for some constant
$a$, is suggested by the request that there be no overlapping of the
resonant domains (of width $O(\epsilon^{1/2})$) formed around each
of the resonant manifolds $k\cdot\omega(J)=0$, with $|k|\leq K$,
other than the overlapping near the loci at which the manifolds
intersect each other (see Morbidelli 2002, p.119). Fixing $K$ and
requesting that only the harmonics with $|k|\leq K$ be normalized
ensures that divisors $k\cdot \omega(J)$ with $|k|>K$ will not
appear in the series. Furthermore, the domain ${\cal W}_a$ is
crossed by only a {\it finite} number of resonant manifolds with
$|k|\leq K$. More precisely, denoting by $\cal{R}_{\alpha}$ the
subset of integer vectors $k$ for which the resonant manifolds
$k\cdot \omega(J)=0$ intersect the domain $W_{\alpha}$, and by
$b_{\cal{R}_{\alpha}}$ a subset of $\cal{R}_{\alpha}$ such that any
vector of $\cal{R}_{\alpha}$ can be produced by a linear combination
of the vectors of $b_{\cal{R}_{\alpha}}$ with integer coefficients,
the relevant property is that $b_{\cal{R}_{\alpha}}$ contains only a
finite number of vectors $k$ with $|k|\leq K$ and an infinite number
of vectors with $|k|> K$. Denoting by $b^{\leq
K}_{\cal{R}_{\alpha}}$ the subset of vectors $k\in
b_{\cal{R}_{\alpha}}$ with $|k|\leq K$, the resonant module ${\cal
M}_{\alpha,\leq K}$ is determined by:
\begin{equation}\label{resmod}
{\cal M}_{\alpha,\leq K}=\mbox{span}\{b^{\leq K}_{\cal{R}_{\alpha}}\}~~.
\end{equation}
Let ${\cal M}_{\alpha,>K}$ be the complement of ${\cal M}_{\alpha,
\leq K}$ with respect to $\cal{R}_{\alpha}$. Under the `book-keeping' choice
$\lambda\equiv\epsilon$, one then has the following iterative algorithm
of determination of the normal form locally, i.e., within the domain
$W_{\alpha}$:

a) Assuming that $(r-1)$ normalization steps were completed, split
$H^{(r-1)}_r$ as $\tilde{H}^{(r-1)}_r + h^{(r-1)}_r$, where $\tilde{H}^{(r-1)}_r$
contains all the terms with $|k|\leq K$ not belonging to ${\cal M}_{\alpha,\leq K}$.

b) Determine the generating function $\chi_r$ by solving
\begin{equation}\label{homor}
\{H_0,\chi_r\} = -\tilde{H}^{(r-1)}_r
\end{equation}

c) Find the new Hamiltonian
\begin{equation}\label{hamr}
H^{(r)}=\exp(\epsilon^rL_{\chi_r})H^{(r-1)} =
Z_0+\epsilon Z_1 + \ldots + \epsilon^r Z_r + H^{(r)}_{res} + H^{(r)}_{nonres}
\end{equation}
in which $Z_j$, $j=0,\ldots,r$ are normal form terms belonging to the
resonant module ${\cal M}_{\alpha,\leq K}$, $H^{(r)}_{res}$ contains all the
Fourier terms belonging to ${\cal M}_{\alpha,>K}$, and $H^{(r)}_{nonres}$
contains the remaining Fourier terms that are not in any of the previously
determined sets. This completes one step of the iteration algorithm. As shown
in Morbidelli and Giorgilli (1997), while all the terms of order higher
than $r$ in these two functions are bounded by a quantity $O(\exp(-K\sigma))$,
i.e., they are both exponentially small in $K$, the terms of $H^{(r)}_{res}$
are the ones making the leading contribution to the remainder.

\subsection{A modified algorithm}

Two problems arise in the implementation of the above algorithm:

a) Eqs.(\ref{homosol1}) or (\ref{homosol2}) imply that the denominators of
all the series coefficients are functions of the actions, in the form of
products of divisors $k\cdot\omega(J)$. This implies that even if $H_0(J)$ is
polynomial in the actions, the Fourier coefficients of the terms
$\exp(ik\cdot\phi)$ in $H^{(r)}$ or $\chi^{(r)}$ are in general rational
in the actions. The extra computational load of dealing with rational
expressions (and their derivatives entering into Poisson brackets) makes
the whole algorithm hardly tractable in this form. We defer the solution
of this problem after the analysis of point (b) below, to which it is linked.

b) One needs to store many more intermediate coefficients than those
eventually needed in order to have a complete knowledge of the
transformed Hamiltonian or of the generating function within any
specified domain of interest. This is a rather technical problem,
the solution of which is, nevertheless, crucial in constructing an
efficient algorithm for the computation of the normal form. A
detailed quantitative discussion of this problem is deferred to the
appendix. Here we state the main result: Under the book-keeping
scheme $\lambda\equiv\epsilon$, assume we are interested in
specifying the transformed Hamiltonian in a `domain of interest'
defined by the maximum orders $(r_{max},K_{max})$, where $r_{max}$
denotes the maximum order in the book-keeping variable $\epsilon$
and $K_{max}$ denotes the maximum order in Fourier space. Then, in
order to be able to specify all the terms belonging to the `domain
of interest', we must store, at intermediate normalization orders
$r\leq r_{max}$, all the Hamiltonian terms of Fourier order $|k|\leq
K(r_{max}-r) +K_{max}$. As shown in the appendix these intermediate
terms outnumber by far the terms finally stored within the `domain
of interest'.

Issue (b) can be resolved by introducing a different book-keeping of
the terms, on the basis of their {\it Fourier order} rather than
$\epsilon-$order. This organization of the series is analyzed in
detail in Giorgilli (2002, pp.86-87) and we present the main points
of this analysis in the appendix. Essentially, it reflects the fact
that for any order $r\leq r_{max}$, many Fourier terms with a
coefficient $\epsilon^r$ in front, which would thus be formally
stored as of order $\lambda^r$ under the book-keeping
$\epsilon\equiv\lambda$, are actually of much smaller size than
$\epsilon^r$, because the size of any $\exp(ik\cdot\phi)$ term
scales as $(e^{-\sigma})^{|k|}$ by virtue of the analyticity
condition. This scaling is already present in the original
Hamiltonian and it propagates at all subsequent normalization steps.
Precisely, this fact is recognized by `book-keeping' the terms on
the basis of their Fourier rather than $\epsilon$-order. The exact
algorithm, which replaces the algorithm of subsection (2.3), reads
as follows:

1) Define $K'=\max\{[1/\sigma],1\}$.

2) Place a book-keeping coefficient $\lambda^p$ in front of each Fourier
term of the form $\exp(ik\cdot\phi)$ in $H_1$, where $p=[|k|/K']+1$.

3) Carry on the normalization (Eqs.(\ref{nfrec})) by successive powers
of $\lambda$.

As shown in the appendix, with such an algorithm there are no extra
terms, outside the domain of interest, that need to be computed.

Furthermore, it is possible to incorporate a solution to issue (a)
in the same algorithm. First, we notice that, as defined in step (1)
above, the constant $K'$ does not pose an upper bound (like $K$) in
the order of the Fourier harmonics being normalized at successive
steps, since the algorithm requests that terms of increasing order
$rK'$ be normalized at the r-th normalization step. In principle,
this would cause a problem when $r$ becomes larger than $K/K'$. But
in practice, one wishes to avoid the problem of the appearance of
the actions in the denominators of the Fourier coefficients for both
$|k|\leq K$ or $|k|>K$, that is for both $r\leq K/K'$ and $r>K/K'$.
It can be shown (section 3) that if the domain ${\cal W}_a$ is
'resonant', that is ${\cal M}_{\alpha,\leq K}\neq \{0,0,\ldots,0\}$,
the width of ${\cal W}_a$ scales with $\epsilon$ as
$O(\epsilon^{1/2})$. Non-resonant domains can also be covered by
subdomains of size $O(\epsilon^{1/2})$. Choosing a point $J_*$ in
the interior of the domain ${\cal W}_{\alpha}$, we then develop the
Hamiltonian locally, within ${\cal W}_{\alpha}$, by expanding the
actions as
\begin{equation}\label{newact}
J=J_*+\epsilon^{1/2}I~~.
\end{equation}
The transformation $J\rightarrow I$ is not canonical under the usual
Poisson structure, but the dynamics remains unaltered if a new Hamiltonian
\begin{equation}\label{hamsc}
H'=\epsilon^{-1/2}H
\end{equation}
substitutes $H$ as generator of the equations of motion. For example,
in a Thirring-type model
\begin{equation}\label{ham2d}
H={J_1^2+J_2^2\over 2} + \epsilon\sum_{k_1,k_2}
c_{k_1,k_2} \exp(i(k_1\phi_1+k_2\phi_2))
\end{equation}
we get, introducing also the `book-keeping' parameter $\lambda$:
\begin{eqnarray}\label{ham2dexp}
H'&=&const+J_{1*}I_1+J_{2*}I_2
+\epsilon^{1/2}\bigg(\lambda{I_1^2+I_2^2\over 2} \nonumber\\
&+&\sum_{k_1,k_2} \lambda^{[|k|/K']+1}c_{k_1,k_2} \exp(i(k_1\phi_1+k_2\phi_2))\bigg)~~.
\end{eqnarray}
Consequently, the perturbation has been rescaled to
$\epsilon^{1/2}$, i.e., it follows the size $O(\epsilon^{1/2})$ of
the domain ${\cal W}_a$. Nevertheless, the unperturbed frequencies
are now $\omega(J_*)\equiv\omega_*$, thus they do not depend on the
actions, since the terms quadratic in the actions are now formally
of order $\lambda$. This implies that the divisors are of the form
$k\cdot\omega_*$, i.e., independent of the actions. Consequently,
the algorithm proceeds with polynomial rather than rational (in the
actions) Fourier coefficients at every step. This also means that
there is no longer need for introducing a K-truncated  resonant
module. Thus, in all calculations the resonant module is specified
only by the resonances between the frequencies $\omega_*$. If
$\omega_*$ are incommensurable, one may specify a module
corresponding either to no resonance or to an approximate resonance
between the frequencies $\omega_*$. As discussed by Morbidelli
(2002, p.48-49), such a construction is still consistent with the
appearance of an exponentially small remainder at the optimal order
of normalization. This is because, as demonstrated in the appendix,
the order of normalization has a linear relation with the maximum
Fourier order of the terms contained in the normal form.
Furthermore, we find that $r_{opt}\sim \epsilon^{-1/2}$ which also
implies $|k|_{opt}\sim\epsilon^{-1/2}$. Hence the dominant terms in
the remainder are of Fourier order $|k|>|k|_{opt}$, i.e., they have
a size $O(\exp(-\sigma\epsilon^{-1/2}))$. This means that
$|k|_{opt}$ plays now a role similar to $K$ in subsection 2.3. It is
interesting to note, however, that such a result is obtained here
{\it a posteriori}, i.e., after the construction of the series. That
is, with the present algorithm one may proceed by calculating the
series up to any desired value of $r$, and then by checking whether
the optimal order $r_{opt}$ was reached. This will also specify
automatically the optimal order $|k|_{opt}$. On the contrary,
following the algorithm of subsection (2.3), one would have to fix a
value of $K$ in advance and then check whether such a value yields
the optimal estimate for the remainder. Otherwise, the calculation
should be repeated from the start, by making a different choice of
$K$.

%%%%%%%%%%%%%%%%%%%%%%%%%%%%%%%%%%%%%%%%%%%%%%%%%%%%%%%%%%%%%%%%%
\section{Results}
%%%%%%%%%%%%%%%%%%%%%%%%%%%%%%%%%%%%%%%%%%%%%%%%%%%%%%%%%%%%%%%%%
%----------------------------------------------------------
\subsection{Hamiltonian model and resonant dynamics}
%----------------------------------------------------------
The Hamiltonian model of Froeschl\'{e} et al. (2000) reads:
\begin{equation}\label{hamfr}
H = H_0+\epsilon H_1 = {I_1^2+I_2^2\over 2}+I_3 +
{\epsilon\over 4+\cos\phi_1+\cos\phi_2+\cos\phi_3}
\end{equation}
where $(I_i,\phi_i)$, $i=1,2,3$ are action - angle variables and
$\epsilon$ is a perturbation parameter (note the change of notation
for the actions, $J\rightarrow I$, with respect to the previous
section, for consistency with the notation of Froeschl\'{e} et al.
(2000)).

%_____________________________________________________________
\begin{figure}
\centering
\includegraphics[scale=0.3]{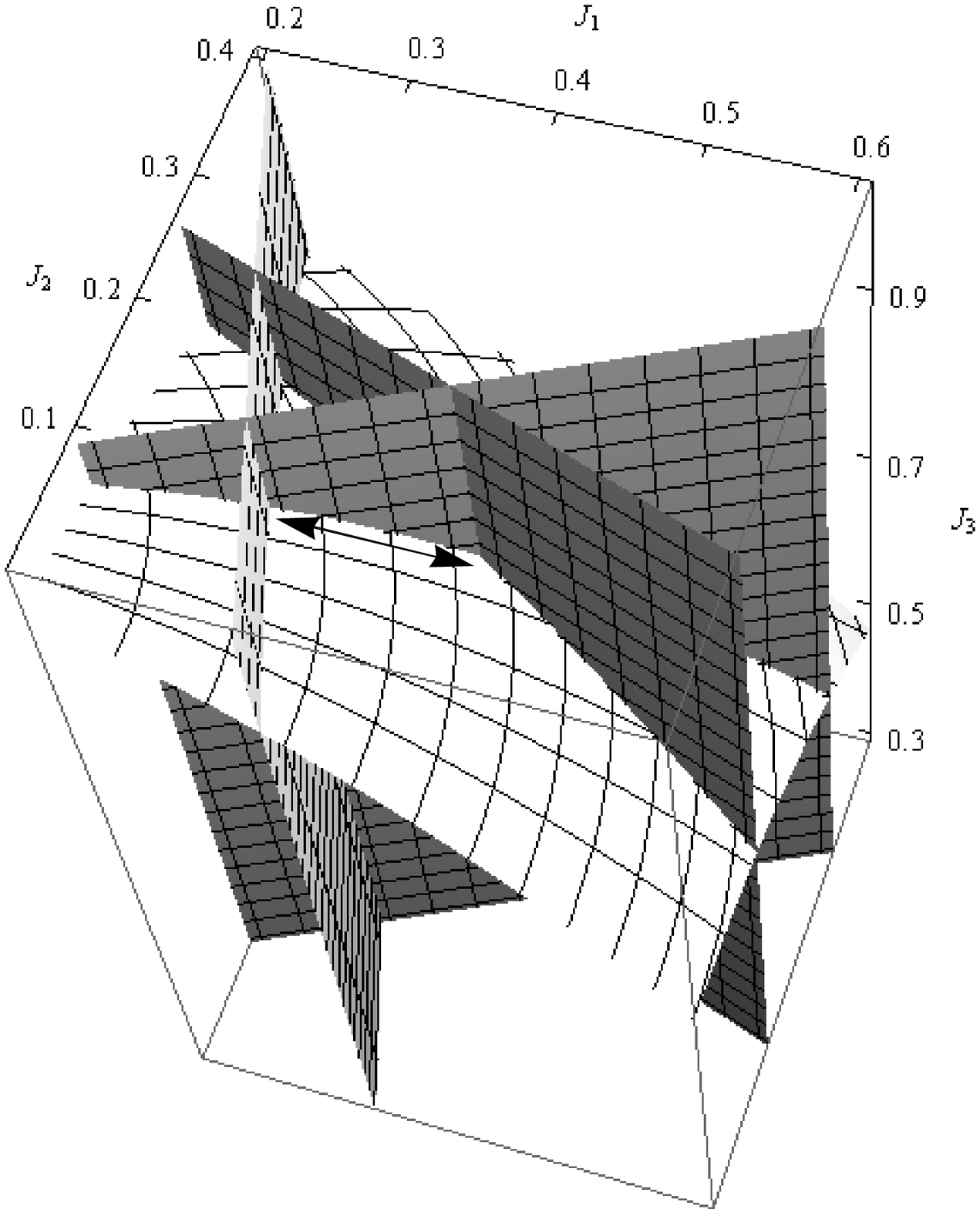}
\includegraphics[scale=0.45]{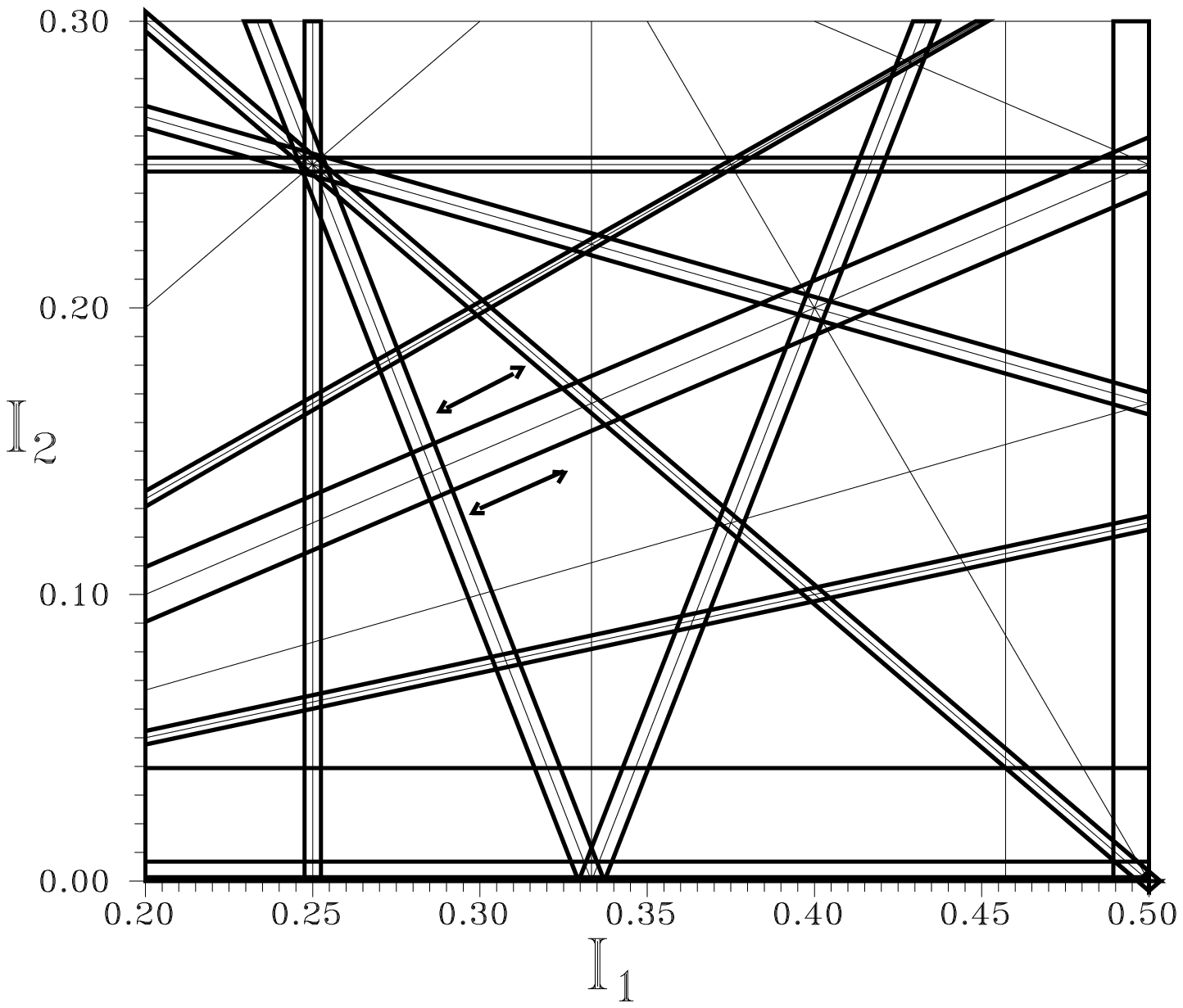}
\caption{(a) The white grid-lined surface is a part of the
paraboloid of constant energy $I_1^2+I_2^2+2I_3=2E$ in the action
space for the value of the energy $E=1$. The three gray-shaded
planes correspond to the resonances $I_1-2I_2=0$ (parallel to the
double arrow), $3I_1+I_2-1=0$, (to the left of the double arrow),
and $2I_1+2I_2-1=0$ (to the right of the double arrow). The
intersection of all the resonant planes with the paraboloid of
constant energy produces a set of parabolic curves which is the 'web
of resonances'. (b) The projection of the web of resonances $k_1I_1
+k_2I_2+k_3=0$ for $|k|\leq 5$ on the $(I_1,I_2)$ plane. Whenever
the coefficient of the term (for one particular resonant vector $k$)
of the Fourier development of $H_1$ is positive, we only plot a
single line corresponding to the projection of the associated
parabolic curve on the plane $(I_1,I_2)$. If this coefficient is
negative, we plot the central line and two other parallel lines
marking the borders of the resonance as obtained by an analytic
estimate of the associated separatrix width when $\epsilon=0.003$.
These rules correspond to a Poincar\'{e} surface of section
$\phi_R=0$, where $\phi_R$ is the resonant angle associated with
each resonance (see text for details), and due to this reason Fig.1b
is directly comparable to Fig.1 of Lega et al. (2003). The analysis
in that paper and below refers to orbits exhibiting chaotic
diffusion along the direction indicated by the double arrows at the
borders of the resonance $I_1-2I_2=0$.} \label{}
\end{figure}
%_____________________________________________________________
The integrable part $H_0$ produces a simple dynamics: $\dot{I_i}=0$
and $\dot{\phi}_1=\omega_{0,1}=I_1$,
$\dot{\phi}_2=\omega_{0,2}=I_2$, $\dot{\phi}_3=\omega_{0,3}=1$. Thus
all three actions are integrals of motion and all three angles
circulate with constant frequencies. The surface of constant energy
$H=E$ is a paraboloid in the action space given by
$I_1^2+I_2^2+2I_3=2E$ (Fig.~1a). On the other hand, the resonances
$k_1\omega_{0,1}+k_2\omega_{0,2} +k_3\omega_{0,3}=$
$k_1I_1+k_2I_2+k_3=0$, with $k\equiv(k_1,k_2,k_3)\in \mathbb{Z}^3$,
$|k|\equiv|k_1|+|k_2|+|k_3|\neq 0$, define planes in the action
space which are always normal to the plane $(I_1,I_2)$. The
intersections of the resonant planes with the surface of constant
energy define the web of resonances on this surface which is a set
of parabolic curves (Fig.~1a). When viewed from the top of Fig.1a,
the projection of these curves on the plane $(I_1,I_2)$ defines a
set of straight lines (Fig.~1b).

When $\epsilon=0$ all the points on the surface of constant energy
of Fig.1a are neutral equilibria, which correspond to invariant
3-tori in the six-dimensional phase space. If, however,
$\epsilon\neq 0$, an $O(\epsilon)$ volume of invariant tori of the
phase space are destroyed, according to the KAM theorem, and
replaced by a chaotic subset of orbits. The constant energy
condition foliates the phase space into 5-dimensional hypersurfaces.
For any fixed value of the angles, the projection of one
hypersurface on the space of actions defines a manifold resembling
to a paraboloid, like in Fig.~1a, which, however, has some
deformation of order $O(\epsilon)$. The chaotic orbits can drift on
this manifold. We shall focus on orbits sliding on the manifold of
constant energy along the chaotic borders of resonances in the
direction indicated by the arrows of Figs.1a,b. As shown below, the
planes of resonances $k\cdot(I_1,I_2,1)=0$ of the unperturbed system
are transformed to resonant manifolds of the perturbed system that
also resemble to planes, but with some thickness of order
$O(|h_k\epsilon|^{1/2}))$, where $h_k$ is the coefficient of the
term $\exp(ik\cdot\phi)$ in the Fourier development of the
perturbation $H_1$. Thus the intersections of the resonant manifolds
with the manifold of constant energy form resonant zones of
thickness $O(|h_k\epsilon|^{1/2})$. When projected on the plane
$(I_1,I_2)$ the web of resonances looks like in Fig.1b (the reason
why in some cases we show one single line for one resonance while in
other cases we show a triple line is discussed below). The chaotic
orbits studied by Lega et al. (2003) are calculated at a value of
$\epsilon$ at which the system is in the so-called `Nekhoroshev
regime', i.e. the width of the resonant zones is small and there is
no significant resonance overlap. The orbits drift slowly along the
border of the resonance $(k_1,k_2,k_3)=(1,-2,0)$, starting with
initial conditions close to, but outside the intersection of this
resonance with a different resonance, namely
$(k_1,k_2,k_3)=(3,1,-1)$. The chaotic drift along the zone of the
resonance $(1,-2,0)$ produces a  slow secular change of the value of
the action $I_F= 2I_1+I_2$ (see Eqs.(\ref{genres} -- \ref{hampen})
below), which is the action associated with the oscillation in the
so-called ``direction of fast drift'', i.e., normal to the resonant
plane. The change of the value of $I_F$ with time was used by Lega
et al. (2003) in order to measure numerically the coefficient of the
chaotic diffusion. The latter causes also a slow change of the value
of the action $I_3$, due to the confinement of the orbits on the
manifold of constant energy.

%_____________________________________________________________
\begin{figure}
\centering
\includegraphics[scale=0.55]{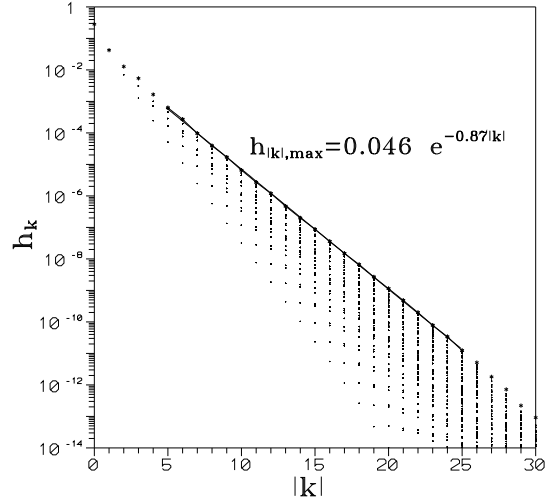}
\caption{The moduli of the coefficients $h_k$ of the Fourier development
(\ref{fourier}) as a function of the modulus $|k|$ for all resonant
terms up to $|k|=40$. The straight solid line is an exponential fit
to the maxima of the $|h_k|$ values as a function of $|k|$. }
\label{}
\end{figure}
%_____________________________________________________________
An estimate of the width of resonances can be made as follows. If we
complexify the angles in a domain of $\mathbb{C}$ containing the
real 3-torus, i.e., if we make the replacement $\phi_j\rightarrow
\phi_j+iu_j$ in $H_1$, with $\phi_j,u_j$ real, then in any direction
of the space $(u_1,u_2,u_3)$ the quantities $u_j$ can vary from zero
up to a value at which $H_1$ becomes singular. However, there is a
lower bound on the distance of all the singularities of $H_1$ from
the real 3-torus, i.e., from the value $(u_1,u_2,u_3)=(0,0,0)$.
Thus, the values of $u_j$ at the point of the closest singularity
cannot be all infinitesimally small. In fact, if we develop the
denominator of $H_1$ up to terms of second degree in $u_j$, we find
an estimate of the position of the singularities of $H_1$ close to
the real 3-torus given by the values $u_j=u_{j,s}$ satisfying
$$
u_{1,s}^2\cos\phi_1+u_{2,s}^2\cos\phi_2+u_{3,s}^2\cos\phi_3
+8+2(\cos\phi_1+\cos\phi_2+\cos\phi_3)=0
$$
and
$$
u_{1,s}\sin\phi_1+u_{2,s}\sin\phi_2+u_{3,s}\sin\phi_3=0~~.
$$
The first of the above equalities implies
$$
u_{1,s}^2+u_{2,s}^2+u_{3,s}^2\geq
|u_{1,s}^2\cos\phi_1+u_{2,s}^2\cos\phi_2+u_{3,s}^2\cos\phi_3| \geq
2~~.
$$
The interior of any parallelepiped in the u-space defined by the
three inequalities $|u_j|<\sigma_j$, for some positive constant
numbers $\sigma_j$, $j=1,2,3$, constitutes a domain of analyticity
of $H_1$ if it is contained within the sphere
$u_1^2+u_2^2+u_3^2=\sqrt{2}$. The optimal estimate for the domain of
analyticity corresponds to the parallelepiped with maximum volume
$8\sigma_1\sigma_2\sigma_3$,  that is to the choice
$\sigma_1=\sigma_2=\sigma_3=\sqrt{2/3}=0.82..\simeq\sigma$.

On the other hand, a more precise value of $\sigma$ can be found by
expressing $H_1$ as a triple Fourier series
\begin{equation}\label{fourier}
{1\over 4+\cos\phi_1+\cos\phi_2+\cos\phi_3} =
\sum_{k_1=-\infty}^{\infty}
\sum_{k_2=-\infty}^{\infty}
\sum_{k_3=-\infty}^{\infty}
h_k\exp(ik\cdot\phi)
\end{equation}
where $k\equiv(k_1,k_2,k_3)$, $\phi\equiv(\phi_1,\phi_2,\phi_3)$. According to
the Fourier theorem on analytic functions, the coefficients $|h_k|$ are bounded
from above by
\begin{equation}\label{expdecay}
|h_k|\leq A\exp(-|k|\sigma)
\end{equation}
with $|k|=|k_1|+|k_2|+|k_3|$ and $A,\sigma$ positive constants. Thus
Eq.(\ref{expdecay}) predicts that the coefficients $h_k$ decay
exponentially. This is shown in Fig.~2, where we plot the values
$|h_k|$, calculated by a series development of $H_1$ using
Mathematica up to the 40th order, against $|k|$. A numerical fitting
to the upper bound of this diagram (solid line) yields $A=0.046$,
and $\sigma = 0.87$. The latter value is close to the estimate
$\sigma=\sqrt{2/3}=0.82$ found above heuristically.

Having determined the size of the Fourier coefficients $h_k$, the normal
form construction described in subsection (2.4) eliminates all harmonics
in the Hamiltonian perturbation $H_1$ (written in the form (\ref{fourier})),
except for the harmonic $h_k\exp(ik\cdot\phi)$ corresponding to the particular
resonance into consideration. To the lowest order, the resonant normal form reads:
\begin{equation}\label{resnf0}
H_{res} = {I_1^2+I_2^2\over 2} + I_3 +2\epsilon h_k\cos(k\cdot\phi)+...~~,
\end{equation}
where we have used the property $h_k=h_{-k}$ following from the
reality and even parity with respect to the angles $\phi_i$ of the
Hamiltonian perturbation $H_1$ . Introducing resonant action - angle
variables $(I_R,\phi_R)$, $(I_F,\phi_F)$ via the generating
function:
\begin{equation}\label{genres}
S=(k\cdot\phi)\big(I_R-{k_3\over k_1^2+k_2^2}\big)
+(m_1\phi_1+m_2\phi_2)I_F+\phi_3I'_3
\end{equation}
where the integers $m_1,m_2$ are any pair satisfying $m_1k_1+m_2k_2=0$, the
resulting canonical transformation is
\begin{eqnarray}\label{restran}
I_1&=&k_1\big(I_R-{k_3\over k_1^2+k_2^2}\big)+m_1I_F,
~~I_2=k_2\big(I_R-{k_3\over k_1^2+k_2^2}\big)+m_2I_F,\nonumber \\
\phi_R&=&k_1\phi_1+k_2\phi_2+k_3\phi_3,~~~~~~~~~\phi_F=m_1\phi_1+m_2\phi_2,\\
& &~~~~~~~~~~~~~~~~~~~I_3=I_3',~~\phi_3'=\phi_3\nonumber~~.
\end{eqnarray}
By virtue of (\ref{restran}), the resonant Hamiltonian (\ref{resnf0}) takes
the form (apart from a constant):
\begin{equation}\label{hampen}
H_{res} = {1\over 2}(m_1^2+m_2^2)I_F^2+I_3+{1\over
2}(k_1^2+k_2^2)I_R^2 +2\epsilon h_k\cos\phi_R
\end{equation}
i.e., it is split in two parts depending only on the actions $I_F$, $I_3$,
and a third part which is the pendulum Hamiltonian:
\begin{equation}\label{hampen2}
H_{pend}={1\over 2}(k_1^2+k_2^2)I_R^2 +2\epsilon h_k\cos\phi_R~~.
\end{equation}
The actions $I_3$ and $I_F=(m_1I_1+m_2I_2)/(m_1^2+m_2^2)$ are
integrals of the Hamiltonian $H_{res}$ (in the case of the resonance
$(k_1,k_2,k_3)=(1,-2,0)$ we have $m_1=2,m_2=1$ and the precise
definition of the fast action is $I_F=(2I_1+I_2)/5$ which differs by
a factor $1/5$ from the definition given at the beginning of this
section for that particular resonance). On the other hand, the width
of the resonance is determined by the separatrix half-width of the
pendulum Hamiltonian $H_{pend}$
\begin{equation}\label{sepwidth}
\Delta I_{R,sep} = \sqrt{{8|h_k\epsilon|\over k_1^2+k_2^2}}~~.
\end{equation}
The phase portrait of $H_{pend}$ is shown schematically in Fig.~3.
In reality, the ideal separatrix given by (\ref{hampen2}) should be
replaced by a thin chaotic layer produced by the weakly chaotic
motion near the separatrix due to to higher order coupling terms of
the Hamiltonian. When $\epsilon$ is small, however, the width of the
chaotic zones is very small. In that case, Eq.(\ref{sepwidth}) can
be used to approximate the maximum normal distance to the resonance
between the upper and lower branches of the separatrix-like chaotic
zones.

The resonant action $I_R$ changes in time according to
$$
\dot{I}_R = -\partial H_{pend}/\partial\phi_R = \epsilon h_k\sin\phi_R~~.
$$
Any variation $\Delta I_R$ of the resonant action results in variations
$\Delta I_1$, $\Delta I_2$, such as to respect the integral $I_F$, i.e.,
$\Delta I_F=0$. Eq.(\ref{restran}) then gives
\begin{equation}\label{di}
\Delta I_1 = k_1\Delta I_R,~~~\Delta I_2 = k_2\Delta I_R
\end{equation}

%_____________________________________________________________
\begin{figure}
\centering
\includegraphics[scale=0.5]{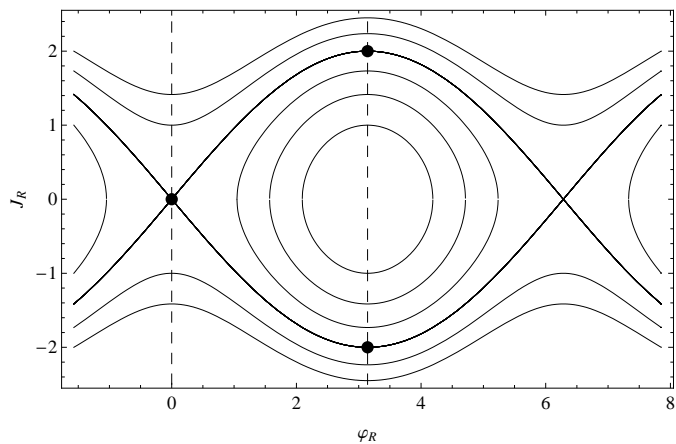}
\caption{The pendulum phase portrait for $H_{pend}$ given by Eq.(\ref{hampen2})
(arbitrary units). The left and right dashed vertical lines indicate the section
$\phi_R=0$ when $h_k>0$ and $h_k<0$ respectively (i.e. $\varphi_R=\phi_R$ and
$\varphi_R=\phi_R-\pi$ respectively). }
\label{}
\end{figure}
%_____________________________________________________________
When the resonance web is projected on the $(I_1,I_2)$ plane, as in
Fig.~1b, the variations $\Delta I_1$ and $\Delta I_2$ of
Eq.(\ref{di}) correspond to motions {\it across} the resonance,
i.e., along a line normal to the resonance line
$k_1I_1+k_2I_2+k_3=0$ (this is called the ``direction of the fast
drift'' by Froeschl\'{e} et al. (2000)) On the other hand, when the
web of resonances is visualized numerically in the action space, as,
for example, by the method of the Fast Lyapunov indicator
(Froeschl\'{e} et al. 2000), one has to make an appropriate choice
of Poincar\'{e} surface of section in order to eliminate the angles
from the calculation. The choice $\phi_3=0$, $|\phi_1|+|\phi_2|\leq
0.05$ made by these authors corresponds essentially to setting the
resonant angle $\phi_R=k_1\phi_1 +k_2\phi_2+k_3\phi_3$ to a value
very close to zero, i.e., $\phi_R\approx 0$. We then see that this
means to plot (a) two sets of points in the plane $(I_1,I_2)$,
corresponding to passing close to the maxima or minima of the
theoretical separatrices, when $h_k<0$, or (b) one set of points on
the plane $(I_1,I_2)$, corresponding to passing close to the X-point
of each separatrix, when $h_k>0$. Consequently, the thin chaotic
borders of the resonances will appear as follows on the surface of
section: In case (a) ($h_k<0$) we see two lines parallel to the
resonant line $k_1I_1+k_2I_2+k_3=0$. These lines define a zone of
width determined by Eq.(\ref{di}). In case (b) ($h_k>0$) we see only
one line coinciding with the resonance line itself. These rules are
followed in the plot of Fig.1b, for all the resonances with $|k|\leq
5$. This figure is to be compared with Fig.1 of Lega et al. (2003).
The two figures compare well not only qualitatively, but also
quantitatively, i.e,, the theoretical resonance widths found above
are very close to the widths determined numerically by the FLI
method.

%--------------------------------------------------------------------
\subsection{Normal form and Nekhoroshev estimates}
%--------------------------------------------------------------------
The resonance dealt with by Lega et al. (2003) is $I_1-2I_2=0$. The diffusion
takes place on the thin chaotic border along this resonance (Fig.2 of Lega et
al. (2003)) when the initial conditions are taken in a small region of the thin
chaotic border. In order to perform a transformation of the type given by
Eqs.(\ref{newact}) and (\ref{hamsc}), we have to specify the central values
$(I_{1*},I_{2*})\equiv(\omega_{1*},\omega_{2*})$ with respect to which the
Hamiltonian is expanded. By visual inspection of Fig.2 of Lega et al.(2003)
a convenient choice compromising the central values of $I_{i*}$ in all the
panels is $I_{1*}=0.31$, $I_{2*}=0.155$. Renaming the variables of Eq.(\ref{hamfr})
by the same symbols according to $\epsilon^{-1/2}H\rightarrow H$, $\epsilon^{-1/2}
(I_i-I_{i*})\rightarrow I_i$, $i=1,2,3$, and Fourier-expanding up to order 40,
the Hamiltonian reads (apart from a constant):
\begin{eqnarray}\label{hamfrexp}
H &=&H_{0*}(I)+\epsilon^{1/2}H_{1*}(I,\phi)=
\omega_{1*}I_1+\omega_{2*}I_2+I_3+ \nonumber\\
& &\epsilon^{1/2}\bigg(
{I_1^2+I_2^2\over 2}+\sum_{|k|\leq 40}c_k\exp(ik\cdot\phi)
\bigg)~~.
\end{eqnarray}

According to the definition of $K'$ given in subsection 2.4, the term
$\epsilon^{1/2}(I_1^2+I_2^2)/2$ should be considered of order of smallness
$p=1$. Then, it is convenient to slightly modify the definition of $K'$ so
as to render $e^{-K'\sigma}$ comparable to the factor $\epsilon^{1/2}$.
We thus set:
\begin{equation}\label{kprime}
K'=\bigg[-{\log(\epsilon^{1/2})\over\sigma}\bigg]+1~~.
\end{equation}
The range of values of $\epsilon$ for which Lega et al. (2003)
provide numerical data on the diffusion coefficient is $30\leq
\epsilon^{-1}\leq 1000$. For most values within this range $K'$ is equal
to a constant value $K'=3$. We thus simply set $K'=3$ in all the calculations.
This means that the Fourier modes of $H_{1*}$ with $0\leq |k|\leq 2$ are
book-kept as of order 1, modes with $3\leq |k|\leq 5$ as of order 2, etc.
The terms of $H_{0*}$ are of order zero, i.e., they are the ones to be used
in the kernel operator $\{\cdot,H_{0*}\}$ of the homological equation.

Although the computer program allows the user to carry along the powers
of $\epsilon^{1/2}$ in the expansion, the extra requirement of memory in
order to store the associated exponents of $\epsilon^{1/2}$ is prohibitive.
We thus had to make separate runs of the computation of the normal form for
various numerical values of $\epsilon$. That is, one numerical value
of $\epsilon$ was supplied to the file storing the coefficients of
(\ref{hamfrexp}) for each run. Even so, a memory limit of 2GB was
reached by computing about $5\times 10^7$ complex coefficients
of the Hamiltonian and of the Lie generating function per run.
Such a number of terms corresponds to a truncation $r\leq 15$ in
the normalization order, or $K=3\cdot 15-1=44$ in Fourier space.
Despite these limits, it was in the end possible to observe the asymptotic
properties of the Birkhoff series within the whole range of values of
$\epsilon$ down to a value $\epsilon=0.0001$, which is one order of
magnitude smaller than the value reached in the numerical experiments
of Lega et al. (2003).

%_____________________________________________________________
\begin{figure}
\centering
\includegraphics[scale=0.7]{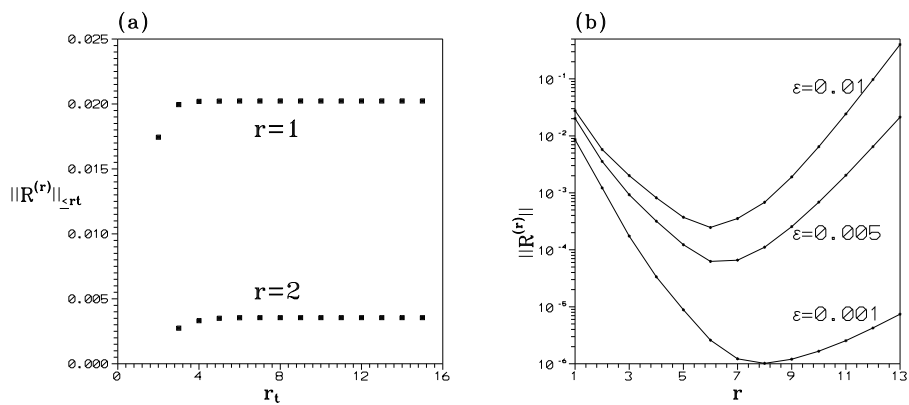}
\caption{(a) The size of the truncated remainder $||R^{(r)}||_{\leq r_t}$
as a function of the truncation order $r_t>r$ after one and after two
normalization steps (upper and lower points respectively) when $\epsilon=0.001$.
(b) The size of the remainder as a function of the order of the normalization
for three different values of $\epsilon$. $||R_{opt}||$ as a function of
$\epsilon$ is given by the size of the remainder at the minimum of each
curve. }
\label{}
\end{figure}
%_____________________________________________________________
After $r$ normalization steps, the Hamiltonian reads:
\begin{equation}\label{nffr}
H^{(r)}=Z^{(r)}(I,\phi)+R^{(r)}_{r+1}++R^{(r)}_{r+2}+...
\end{equation}
The normal form $Z^{(r)}(I,\phi)$ contains all the Fourier terms up to
$|k|\leq 3r-1$ belonging to the resonant module ${\cal M}=\{k: k_1-2k_2=0
~\mbox{and}~k_3=0\}$. The normal form $Z^{(r)}(I,\phi)$ is an integrable
hamiltonian, the integrals, besides the energy, being $I_3$ and $I_F
=2I_1+I_2$. The level lines $I_F=c$ are normal to the resonance line
$I_1-2I_2=0$. Since we are interested in measuring the drift along the
chaotic border of the resonance, we can estimate the values $|I_1|$ and
$|I_2|$ at the chaotic border either numerically, i.e., by measuring the
half-width of the resonant zone in each panel of fig.2 of Lega et al.
(2003), or by using Eq.(\ref{sepwidth}) with
$k_1=1,k_2=-2$. The numerical calculation was obtained by measuring
the half-width $\Delta I$ of the interval along the line $2I_1+I_2=c$
passing through $(I_{1*},I_{2*})=(0.31,0.155)$ with endpoints at the
two sides of the chaotic border marked yellow in Lega et al. (2003),
and by dividing $\Delta I$ by $\epsilon^{1/2}$ according to the rescaling
$(\ref{newact})$. As expected, the estimates for three different values of
$\epsilon$ turned to be approximately equal, yielding average values
$(|I_1|,|I_2|)=(0.0325,0.065)$. These values were inserted in (\ref{nffr})
in order to calculate the norm of the remainder $R^{(r)}(I,\phi)\equiv
R^{(r)}_{r+1}+R^{(r)}_{r+2}+...$. It should be noticed that $I_3$ does
not appear in the remainder because it is a dummy action, i.e.,
it only appears linearly in the unperturbed Hamiltonian.

According to standard theory, the remainder series $R^{(r)}(I,\phi)$
should be convergent. This is shown in Fig.4a, in which the
truncated sum
\begin{equation}\label{remsum}
||R^{(r)}||_{\leq r_t}\equiv\sum_{j=r+1}^{r_t}||R^{(r)}_j||
\end{equation}
is plotted against the truncation order $r_t$ for the remainders of
the first two normalization steps $r=1$ and $r=2$ for
$\epsilon=0.001$ (the norm $||\cdot||$ is taken as the sum of moduli
of all the trigonometric coefficients of the involved function).
Clearly, the remainder shows the tendency to converge to its final
size after summing just its first two or three successive terms.
Furthermore, the size of the remainder, estimated as
$||R^{(r)}||_{\leq 15}$ exhibits the well known behavior expected
for an asymptotic series (Fig.4b). Namely, the size of the remainder
initially decreases, as $r$ increases, up to an optimal order
$r_{opt}(\epsilon)$ corresponding to the minima of the curves of
Fig.4b for different values of $\epsilon$. Beyond the optimal order,
however, the remainder becomes an increasing function of $r$ and one
has $\lim_{r\rightarrow\infty}||R^{(r)}||=\infty$, marking the
eventual divergence of the normalization procedure. From fig.5b we
see clearly that the optimal order $r_{opt}$ increases as $\epsilon$
decreases.

Figures 5 and 6 show now the main result. The abscissa in Fig.5
shows the value of
$||R_{opt}||\equiv||R^{(r_{opt}(\epsilon))}||_{\leq 15}$ for ten
different values of $\epsilon$ in the range $0.001\leq \epsilon\leq
0.02$. The ordinate shows the value of the diffusion coefficient
$D(\epsilon)$ for the same values of $\epsilon$ as given by Lega et
al. (2003). The straight line represents a power-law fitting of the
relation between the optimal remainder and the diffusion
coefficient. The best fit law is $D=1.02 ||R_{opt}||^{2.97}$. Thus,
the scaling is essentially:
\begin{equation}\label{dropt}
D\propto ||R_{opt}||^3~~.
\end{equation}

On the other hand, Fig.6 shows the scaling of $||R_{opt}||$ with
$1/\epsilon$ when $\epsilon$ reaches values as small as
$\epsilon=0.0001$, i.e., one order of magnitude smaller than the
last point of Fig.7 of Lega et al. (2003) (the right vertical dashed
line shows the last point where the numerical calculation of Lega et
al. (2003) was stopped). The two solid lines passing through the
data correspond to a power-law (lower curve) and an exponential law
(upper curve) fitting the data. For small values of $1/\epsilon$
(i.e. for large values of $\epsilon$), the power-law fitting is
better than the exponential fitting. The limit beyond which the
exponential fitting is acceptable is around $\epsilon=0.01$
(indicated by the left vertical dashed line of Fig.6 at
$1/\epsilon=100$). We may identify this limit as roughly
corresponding to the threshold of the Nekhoroshev regime. On the
other hand, in the interval $100\leq 1/\epsilon\leq 1000$, both the
power law and the exponential law yield acceptable fittings.
Nevertheless, beyond the value $1/\epsilon>1000$, the power-law
clearly fails while the exponential fitting follows now narrowly the
data. The numerical best fit exponential law yields
$$
||R_{opt}||\propto\exp(-\epsilon_*/\epsilon^{0.21})~~.
$$

%_____________________________________________________________
\begin{figure}
\centering
\includegraphics[scale=0.55]{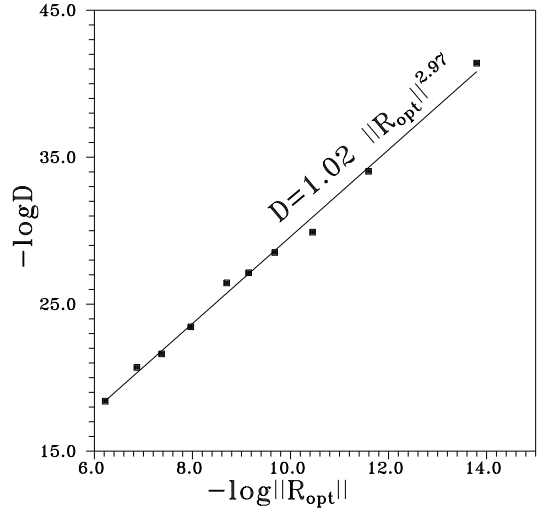}
\caption{The diffusion coefficient $D$ reported in Lega et al.
(2003) as a function of the size of the optimal remainder
$||R_{opt}||$ of the Birkhoff normal form calculated for different
values of $\epsilon$.} \label{}
\end{figure}
%_____________________________________________________________
%_____________________________________________________________
\begin{figure}
\centering
\includegraphics[scale=0.55]{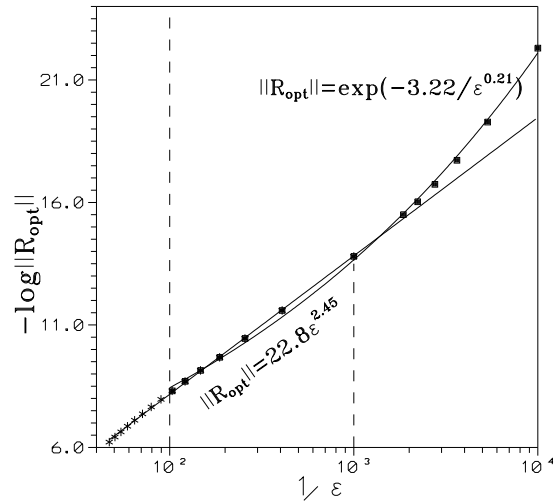}
\caption{The size of the optimal remainder $||R_{opt}||$ as a
function of $1/\epsilon$. The data clearly deviate from a power law
beyond $1/\epsilon=1000$, and they are well fitted by an exponential
law, as predicted by the Nekhoroshev theorem. The left vertical
dashed line gives a lower threshold of $1/\epsilon$ below which the
exponential law is no longer valid. The data are well fitted also by
a power law in the range $10\leq 1/\epsilon\leq 1000$. The right
vertical dashed line shows the last point of the numerical
calculation of the diffusion coefficient by Lega et al. (2003).}
\label{}
\end{figure}
%_____________________________________________________________
We note that Froeschl\'{e} et al. (2005), making numerical
experiments of the diffusion of orbits in a mapping model resembling
to the Hamiltonian (\ref{hamfr}), found an exponential scaling of
the diffusion coefficient with $\epsilon$ in which the best-fit
exponent is $a=0.28$. However, in that paper too the exponential
scaling was unraveled by considering values of $\epsilon$ smaller by
more than one orders of magnitude from a critical threshold value
below which resonances do not significantly overlap. We conclude
that the exponential law predicted by the Nekhoroshev theorem can be
unambiguously seen in the real data by various methods when
$\epsilon$ becomes at least one order of magnitude smaller than the
critical value characterizing the onset of the `Nekhoroshev regime'.

%%%%%%%%%%%%%%%%%%%%%%%%%%%%%%%%%%%%%%
\section{Conclusions and Discussion}
%%%%%%%%%%%%%%%%%%%%%%%%%%%%%%%%%%%%%%

The present paper demonstrates the applicability of the analytical
apparatus of the Nekhoroshev theory in recovering the exponential
scaling of the coefficient of Arnold diffusion
$D=O(\exp(-1/\epsilon))$ along a particular resonance in the model
Hamiltonian system of three degrees of freedom of Froeschl\'{e} et
al. (2000). In particular:

1) The normalization algorithm for the Hamiltonian, based on an
organization of the perturbation series in terms of the Fourier
order of the terms (rather than on powers of $\epsilon$), is
presented, and its benefits are analyzed in detail.

2) The implementation of this algorithm on the computer allowed us
to reach the optimal order of normalization, at which the size of
the remainder $||R_{opt}||$ of the normal form becomes minimal.

3) The coefficient of Arnold diffusion $D$, as determined
numerically by Lega et al. (2003), is found to scale with the size
of the optimal remainder like $D\propto R_{opt}^3$.

4) The size of the optimal remainder is found to scale exponentially
with the inverse of the perturbation parameter, namely
$||R_{opt}||\propto \exp(-1/\epsilon^a)$, with $a\simeq 0.21$. The
exponential scaling clearly shows when one considers values of
$\epsilon$ as small as $\epsilon=10^{-4}$, i.e., one order of
magnitude smaller than in the numerical experiments of Lega et al.
(2003).

Estimates on the speed of Arnold diffusion based on numerical experiments have
been given in the literature by a number of authors. We note in particular the
early work of Kaneko and Konishi (1989), in which an exponential scaling law
is found with $a$ between $0.1$ and $0.3$, i.e., consistent with the present
results.

On the other hand, Wood et al. (1990) provided estimates of the
diffusion coefficient in standard-like multidimensional symplectic
maps on the basis of the Arnold-Melnikov method, i.e., the
exponentially small splitting of separatrices due to the presence of
higher order coupling terms in the resonant Hamiltonian. An
exponential scaling of $D$ with $1/\epsilon$ was also found in that
case, favoring though a value of $a$ rather close to $a=1/2$. Thus,
a detailed comparison of the results by the Nekhoroshev and
Arnold-Melnikov theories is in order. To our knowledge, the only
hint in that direction is a paper by Morbidelli and Giorgilli
(1997). Nevertheless, more work is necessary in order to clarify
this connection by specific quantitative studies, as well as to
compare the predictions of the two theories for Arnold diffusion in
various types of multidimensional systems.

Finally, the fact that the scaling of $||R_{opt}||$ with $\epsilon$ appears
also as a power law in a transient interval of values of $\epsilon$, before
the onset of the exponential regime, is consistent with some past theoretical
work (Chirikov et al. 1985) calling such a transient regime `modulational
diffusion'. According to this theory, there is an intermediate interval of
values of $\epsilon$ within which many high order resonances, located in
the chaotic border of the low-order resonance along which the diffusion
takes place, still overlap. We conjecture that these are resonances of
some order $|k|$ which must be above, but close to a truncation order
$K$ estimated as $K\sim 1/\epsilon^a$ (according to the analysis
of subsection 2.3).

At any rate, the results of the present paper provide a clear relation between
the local value of the diffusion coefficient $D$ along a resonance, on the one
hand, and the size of the optimal remainder $||R_{opt}||$ of the normal form
for the same resonance, on the other hand. Furthermore, for $\epsilon$ sufficiently
small $||R_{opt}||$ is found to scale with $\epsilon$ precisely as predicted
by the Nekhoroshev theory. Thus, the final conclusion of our study that
{\it the analytical techniques of the Nekhoroshev theory can be used
with much profit, in order to construct precise quantitative estimates
of the speed of Arnold diffusion in Hamiltonian systems}.\\
\\
{\bf Acknowledgements:} I would like to thank Professors A.
Giorgilli, for explaining to me the benefits of normalizing a series
according to the Fourier order of the terms, C. Froeschl\'{e}, for
an invitation to Nice where this project started, and Z.
Kne\v{z}evi\'{c} and G. Contopoulos, for many useful comments on the
manuscript. C. Lhotka made a careful reading and prepared Fig.1.

\begin{center}
{\bf Appendix}
\end{center}

The following is a more technical analysis of point (b) of
subsection 2.4 referring to the organization of the perturbation
series in powers of the small parameter $\epsilon$.

Let $r_{max}$ be the maximum order in the book-keeping variable
$\lambda\equiv\epsilon$, $K_{max}$ be the maximum order in Fourier
space specifying a domain in which we are interested in computing
the Hamiltonian, and $K$ the order of truncation in Fourier space
beyond which terms are not normalized (in general we have to set
$K_{max}>K$ so that some remainder terms of the Hamiltonian be also
stored). Figure~7a shows this domain as a gray-shaded parallelogram
in the space $(r,|k|)$. Now, it can be readily seen that many terms
of the Hamiltonian which lie {\it outside} the domain of interest at
some particular order $r$ interact with the generating function in
such a way as to produce, at some subsequent order, terms which lie
{\it inside} the domain of interest. Indeed, consider a term of the
form $\epsilon^{r_1}a(J)\exp(ik\cdot\phi)$ in the generating
function, with $r_1\leq r_{max}$ and $|k|\leq K<K_{max}$ (inside the
domain), and a term $\epsilon^{r_2}a'(J)\exp(ik'\cdot\phi)$ in the
Hamiltonian, with $r_2\leq r_{max}$ and $|k|> K_{max}$ (outside the
domain). Assume that $r_1+r_2\leq r_{max}$. Then, a Poisson bracket
operation between the two terms yields a term
$\epsilon^{r_1+r_2}[a(J)k\cdot
\nabla_Ja'(J)-a'(J)k'\cdot\nabla_Ja(J)]\exp(i(k+k')\cdot\phi)$. If
$|k+k'| \leq K_{max}$ (this is possible because the components of
$k$ and $k'$ are added algebraically), the so produced term lies
inside the domain of interest, despite the fact that the Hamiltonian
term from which it originated lies outside the domain of interest.
%_____________________________________________________________
\begin{figure}
\centering
\includegraphics[scale=0.5]{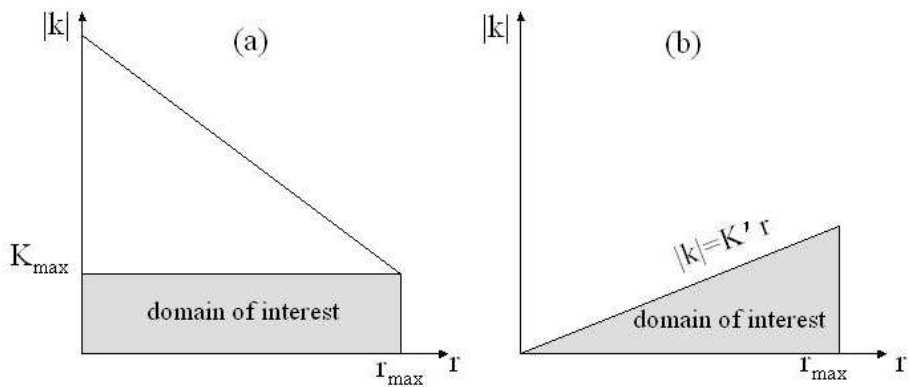}
\caption{(a) The gray parallelogram shows the `domain of interest'
defined as the domain in Fourier space (orders of $|k|$) of the
terms that we wish to store in the transformed Hamiltonian up to an
order $r=r_{max}$ in the book-keeping parameter
$\lambda\equiv\epsilon$. The inclined line, given by
$|k|=K_{max}(r_{max}-r+1)$, represents the upper limit of a larger
domain in Fourier space within which intermediate terms must be
stored at an order $r$, yielding contributions in the gray-shaded
domain at orders larger than $r$. (b) Under an appropriate
book-keeping (see subsection (2.4)) the domain of interest can be
made to coincide with the domain in which terms must be stored. }
\label{}
\end{figure}
%_____________________________________________________________

A careful examination of all the Lie operations taking place up to
the order $r_{max}$ shows that in order to obtain complete knowledge
of the transformed Hamiltonian in a domain $(r_{max},K_{max})$, one
must store, for every order $r\leq r_{max}$, all the Hamiltonian
terms of Fourier order $|k|\leq K(r_{max}-r) +K_{max}$. This
extended domain, shown by the triangular domain of figure~1a,
contains many more terms than those of the domain of interest (the
ratio of the number of terms in the two domains is proportional to
the ratio of the areas of the domains raised to a power $\simeq 2n$,
where $n$ is the number of degrees of freedom).

At this point we may argue that, for sufficiently large $K_{max}$,
the terms with $|k|>K_{max}$ have very small size because of the
analyticity condition (Eq.(\ref{expbd})), so that they can probably
be ignored without great modification of the results. However,
precisely this type of argument brings about the real source of the
above problem, which is the choice of `book-keeping'
$\lambda\equiv\epsilon$. The fact that the triangular domain of
figure~1a contains many more terms than the gray parallelogram in
the same figure simply depicts failure to recognize that for any
order $r\leq r_{max}$, many Fourier terms formally stored as of
order $\lambda^r$ are actually of much smaller size than
$\epsilon^r$. For example, while $H_1$ is an overall $O(1)$
quantity, not all the Fourier terms in $H_1$ have a similar size,
because the size of an $\exp(ik\cdot\phi)$ term in $H_1$ scales as
$(e^{-\sigma})^{|k|}$. This initial scaling propagates also at
subsequent orders of normalization. This fact suggests that the
terms should be `book-kept' differently, i.e., on the basis of their
Fourier order rather than their $\epsilon$-order.

Following the algorithm exposed in subsection 2.4 (steps (1) to
(3)), the domain of interest (figure 1b) is made to coincide with
the domain in which terms should be stored, i.e., terms within the
gray-shaded area can only be produced by Poisson bracket operations
between terms which are also within the gray shaded area. In
particular, the meaning of step (2) is that in a Hamiltonian like
(\ref{ham}) the parameter $\epsilon$ is not so relevant in
characterizing the smallness of the various terms in $H_1$
($\epsilon$ can in fact be incorporated in the book-keeping scheme,
as was done in section 3). A formal analogy can be made with a
polynomial Hamiltonian around an elliptic equilibrium:
\begin{equation}\label{hampol}
H\equiv \sum_{i=1}^n\omega_i{p_i^2+q_i^2\over 2} +
\epsilon\sum_{\sum m_i\geq 3}\alpha_{m_1,m_2,...,m_n}
 q_1^{m_1}q_2^{m_2}...q_n^{m_n}~~.
\end{equation}
In the case of the Hamiltonian (\ref{hampol}) it is customary to
proceed by normalizing iteratively by the degrees of the monomial
terms in the canonical variables rather than by powers of
$\epsilon$. Such a choice underlines that the real smallness in the
polynomial case is the distance $\rho=\sqrt{\sum(q_i^2+p_i^2)}$ from
the equilibrium point. In fact, it is often convenient to rescale
the whole Hamiltonian (\ref{hampol}) so that $\epsilon$ disappears
from the lowest order terms of the polynomial expansion in the
r.h.s. of (\ref{hampol}). The analogy with the generic case becomes
clear by introducing action-angle variables
$q_i=\sqrt{2J_i}\sin\phi_i$, $p_i=\sqrt{2J_i}\cos\phi_i$,
$i=1,\ldots,n$. We then readily find that a Fourier term of order
$\rho^r\sim J^{r/2}$ is always of Fourier order $|k|\leq r$ (and
$|k|$ has the same parity as $r$). The relevant domain of interest
is thus limited by the equation $|k|=r$ which yields a domain very
similar to Fig.7b (the only difference is in the slope of the
limiting upper line, which is $[1/\sigma]$ in Fig.7b and $1$ in the
polynomial case).

\end{article}

\begin{thebibliography}{}

\bibitem{}{}{}
Arnold, V.I., 1964: {\it Sov. Math. Dokl.} {\bf 6}, 581.
\bibitem{}{}{}
Benettin, G., Galgani, L., and Giorgilli, A.: 1985, {\it Cel. Mech.}
{\bf 37}, 1.
\bibitem{}{}{}
Boccaletti, D., and Pucacco, G.: 1996,{\it Theory of Orbits}, Springer, Berlin.
\bibitem{}{}{}
Chirikov, B.V., Lieberman, M.A., Shepelyansky, D.L., and Vivaldi, F.M.:
1985, {\it Physica} {\bf 14D}, 289.
\bibitem{}{}{}
Contopoulos, G.: 2002, {\it Order and Chaos in Dynamical Astronomy}, Springer, Berlin.
\bibitem{}{}{}
Dumas, H.S., and Laskar, J.: 1993, {\it Phys. Rev. Lett.} {\bf 70}, 2975.
\bibitem{}{}{}
Ferraz-Mello, S: 2007, {\it Canonical Perturbation Theories. Degenerate
Systems and Resonance.}, Springer, New York.
\bibitem{}{}{}
Froeschl\'{e}, C., Guzzo, M., and Lega, E.: 2000, {\it Science} {\bf 289}
(5487), 2108.
\bibitem{}{}{}
Froeschl\'{e}, C., Guzzo, M., and Lega, E.: 2005, {\it Cel. Mech. Dyn. Astron.}
{\bf 92}, 243.
\bibitem{}{}{}
Giordano, C.M., and Cincotta, P.M.: 2004, {\it Astron. Astrophys.} {\bf 423}, 745.
\bibitem{}{}{}
Giorgilli, A: 2002, {\it Notes on exponential stability of Hamiltonian systems},
in Dynamical Systems. Part I: Hamiltonian Systems and Celestial Mechanics,
Pubblicazioni della Classe di Scienze, Scuola Normale Superiore, Pisa.
\bibitem{}{}{}
Guzzo, M., Lega, E., and Froeschl\'{e}, C.: 2005, {\it Dis. Con. Dyn. Sys. B}
{\bf 5}, 687.
\bibitem{}{}{}
Kaneko, K., and Konishi, T.: 1989, {\it Phys. Rev. A} {\bf 40}, 6130.
\bibitem{}{}{}
Laskar, J.: 1993, {\it Physica} {\bf D67}, 257.
\bibitem{}{}{}
Lega, E., Guzzo, M., and Froeschl\'{e}, C.: 2003, {\it Physica D} {\bf 182}, 179.
\bibitem{}{}{}
Lochak, P.: 1992, {\it Russ. Math. Surv.} {\bf 47}, 57.
\bibitem{}{}{}
Morbidelli, A.: 2002, {\it Modern Celestial Mechanics. Aspects of Solar System
Dynamics}, Taylor and Francis, London.
\bibitem{}{}{}
Morbidelli, A., and Guzzo, M.: 1997, {\it Cel. Mech. Dyn. Astron.} {\bf 65}, 107.
\bibitem{}{}{}
Morbidelli, A., and Giorgilli, A.: 1997, {\it Physica D} {\bf 102}, 195.
\bibitem{}{}{}
Nekhoroshev, N.N.: 1977, {\it Russ. Math. Surv.} {\bf 32}(6), 1.
\bibitem{}{}{}
P\"{o}shel, J.: 1993, {\it Math. Z.} {\bf 213}, 187.
\bibitem{}{}{}
Skokos, C., Contopoulos, G., and Polymilis, C.: 1997, {\it Cel. Mech. Dyn. Astr.}
{\bf 65}, 223.
\bibitem{}{}{}
Wood, B.P., Lichtenberg, A.J., and Lieberman, M.A.: 1990,{\it Phys.
Rev. A} {\bf 42}, 5885.

\end{thebibliography}
\end{document}